# $C_5H_9N$ Isomers: Pointers to Possible Branched Chain Interstellar Molecules

Emmanuel E. Etim[1,3*], Prasanta Gorai[2], Ankan Das[2], Elangannan Arunan[1]

[1]Inorganic and Physical Chemistry Department, Indian Institute of Science Bangalore, India-560012

[2]Indian Centre for Space Physics, 43 Chalantika, Garia Station Road, Kolkata 700 084, India

[3]Department of Chemical Sciences, Federal University Wukari, Nigeria

*email: emmaetim@gmail.com

**Abstract:** The astronomical observation of isopropyl cyanide further stresses the link between the chemical composition of the ISM and molecular composition of the meteorites in which there is a dominance of branched chain amino acids as compared to the straight. However, observations of more branched chain molecules in ISM will firmly establish this link. In the light of this, we have considered $C_5H_9N$ isomeric group in which the next higher member of the alkyl cyanide and other branched chain isomers belong. High-level quantum chemical calculations have been employed in estimating accurate energies of these isomers. From the results, the only isomer of the group that has been astronomically searched, n-butyl cyanide is not the most stable isomer and therefore, which might explain why its search could only yield upper limits of its column density without a successful detection. Rather, the two most stable isomers of the group are the branched chain isomers; tert-butylnitrile and isobutyl cyanide. Based on the rotational constants of these isomers, it is found that the expected intensity of tert-butylnitrile is the maximum among this isomeric group. Thus, this is proposed as the most probable candidate for astronomical observation. A simple LTE (Local thermodynamic equilibrium) modelling has also been carried out to check the possibility of detecting tert-butyl cyanide in the millimetre-wave region.

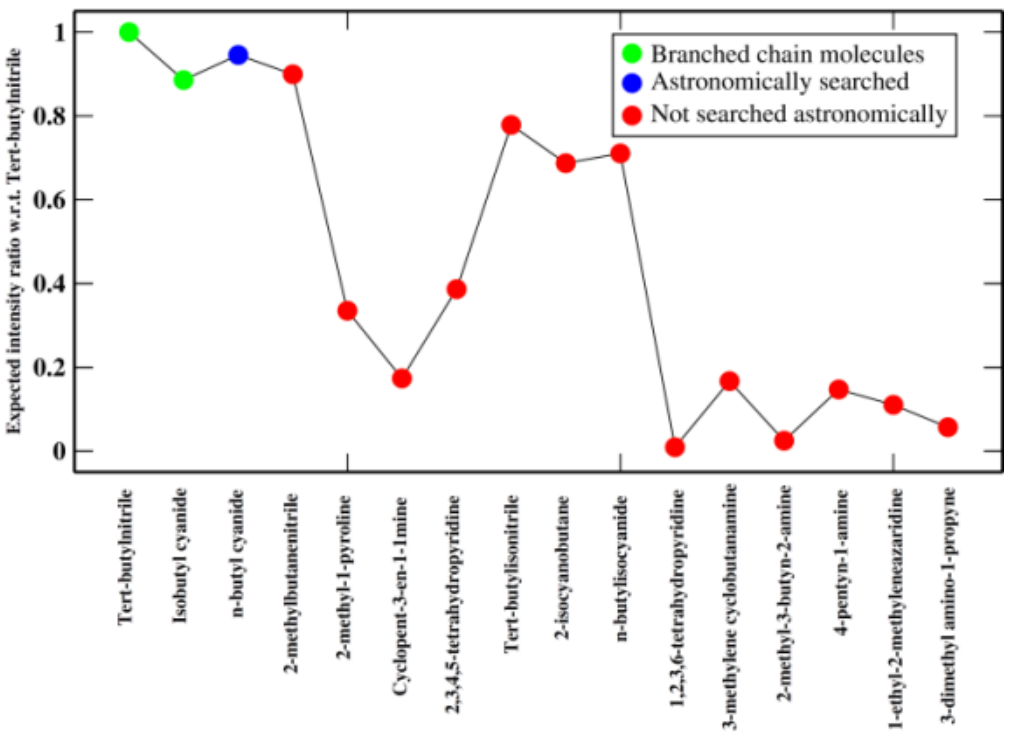

## 1. Introduction

Each unique astronomical observation of a molecule does more than just informing us about its presence, the physics and chemistry from where it was observed rather it also serves as a pointer to other possible molecules in ISM. For example the observations of the O-containing molecules and their corresponding S-analogues do not only re-emphasis the relationship between O and S as elements in the same group rather it can easily be seen that for every known O-containing molecule in ISM, the S-analogue is also likely to be present even when it is yet to be astronomically observed. The positive impact of the recent development and advances in astronomical and spectroscopic techniques is setting the pace for a better understanding of the science (physics and chemistry) of the interstellar space and in probing deep into the interior of the molecular clouds. That about 200 different molecular species have been detected largely via their rotational transitions in the microwave, millimeter-wave, and terahertz region evidently proves a good examination of the interstellar and circumstellar media [1-2]. Explanation of the existence of the observed complex molecules could not be explained without a proper consideration of interstellar dusts [3-8]. These studies proposed that dust grains are actually behaving like catalyst for the formation of complex molecules in space. Various experimental studies also aided this hypothesis [9-10]. The difficulty in detecting molecules in space increases with the complexity of the molecule. From the list of known interstellar and circumstellar molecules, there is a steady decrease in the number of these molecules as the number of atoms increase.

Though different classes and types of molecules have been observed in space, there was no astronomical observation of a branched chain molecule until recently. Isopropyl cyanide was observed in the Sagittarius B2 complex molecular via its rotational transitions [11]. Relative stability of CN containing molecules in astrochemical environments were already been discussed in literatures [12-15]. The astronomical detection of isopropyl cyanide, the first branched molecule in interstellar medium (ISM) came longer than expected being the most stable isomer of the $C_4H_7N$ isomeric group as shown in Table 1 Propyl cyanide; the second most stable isomer of the $C_4H_7N$ isomeric group was observed few years [16] before the branched chain. The delayed observation of isopropyl cyanide could be traced to the fact that its rotational spectrum was only studied to a limited extent in the microwave region before now [11,17]. Table 1 shows the enthalpies of formation for these isomers obtained with the Gaussian 4 (G4) composite method (more information under computational details), the experimental value of this parameter for isopropyl cyanide is indicated with the square bracket in the table. The experimental value [18,19] is in excellent agreement with the calculated value.

**Table 1:** $\Delta_f H^O$ for $C_4H_7N$ isomers and current astronomical status

| Molecule Structure | Molecule | $\Delta_f H^O$ (kcal/mol) | Astronomical status |
|---|---|---|---|
| 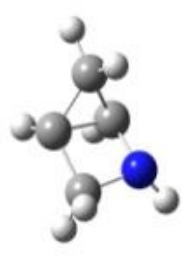 | 2-azabicyclo(2.1.0)pentane | 57.1 | Not observed |
| 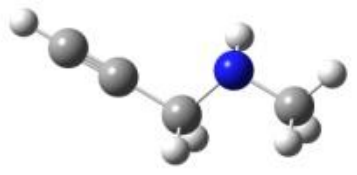 | N-methyl propargylamine | 54.5 | Not observed |
| 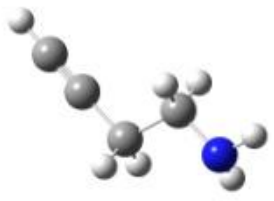 | 3-butyn-1-amine | 48.4 | Not observed |
| 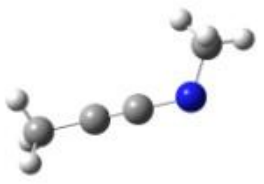 | N-methyl-1-propyn-1-amine | 46.6 | Not observed |
| 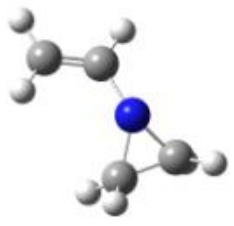 | N-vinylazaridine | 45.9 | Not observed |
| 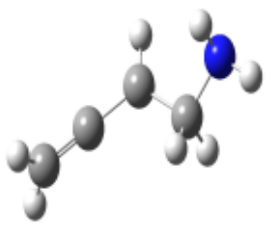 | 2,3-butadiene-1-amine | 43.3 | Not observed |
| 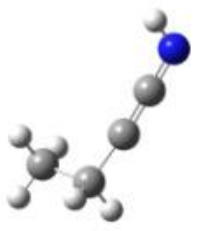 | But-1-en-1-imine | 27.8 | Not observed |
| 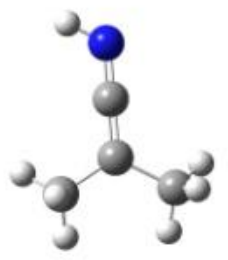 | 2,2-dimethylethylenimine | 25.7 | Not observed |
| 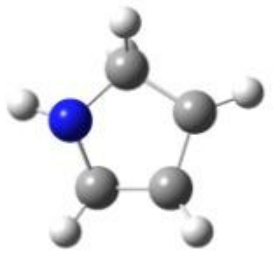 | 3-pyrroline | 25.4 | Not observed |
| 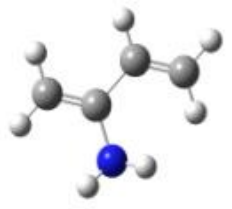 | 2-aminobutadiene | 24.4 | Not observed |
| 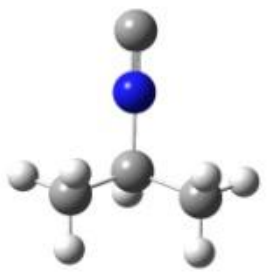 | 2-isocyanopropane | 24.4 | Not observed |
| 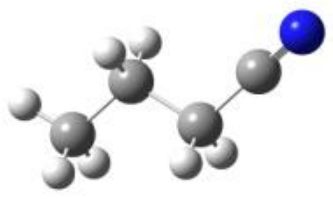 | Propyl cyanide | 5.6 [7.4] | Observed |
| 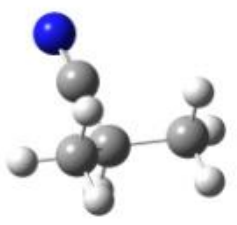 | Isopropyl cyanide | 5.2[5.4] | Observed |

With no amino acid yet to be successfully detected in interstellar or circumstellar medium, about 80 of them have already been identified in meteorites found on earth [20]. The origin of these chemical compounds or their precursors found in meteorites has been traced to the interstellar or circumstellar medium [21], thus suggesting the detection of more complex molecules in the interstellar and circumstellar media than what have been detected so far. This paints a picture of a link between the interstellar medium and the molecular composition of meteorites. Most of the amino acids identified in meteorites are branched amino acids [21-22]. The observation of a branched chain molecule, isopropyl cyanide in the interstellar medium further stresses the possibility of this link between the interstellar medium and the molecular composition of meteorites. However, in order to firmly establish this link, the observation of more branched chain molecules in the interstellar and circumstellar media is required. In the light of this, the $C_5H_9N$ isomeric group, where the next higher member of the alkyl cyanide series belong is examined to see if the branched chain molecules in group could be the possible candidates for astronomical observation as in the case of the $C_4H_7N$ isomeric group.

## 2. Computational Details

All the quantum chemical calculations in this paper were performed with the Gaussian 09 suite of programs [23]. In order to estimate accurate enthalpies of formation for all the molecules in the $C_4H_7N$ (shown in Table 1) and $C_5H_9N$ isomeric groups that are in good agreement with experimental values (where available), the Gaussian 4 (G4) composite method was employed[24-26]. In arriving at an accurate total energy and zero point energy (ZPE) for a given molecule, the G4 composite method performs a sequence of well-defied ab initio molecular calculations [27]. Harmonic frequency calculations revealed all real frequencies confirming that the optimized structures are all minima in the potential energy hypersurface. Rotational and quartic centrifugal-distortional constants values are calculated using B3LYP/ 6-311++G(d,p) method.

## 3. Results and Discussion

In this section, the results of the high-level quantum chemical calculations are presented and discussed. For the four most stable isomers which are good candidates for astronomical searches, information on the available spectroscopic data (from literature) necessary for astronomical observation is given. In Table 2, 16 possible isomers of the $C_5H_9N$ isomeric group with their corresponding enthalpy of formation value ($\Delta_f H^O$), dipole moment and optimized geometry are presented in ascending order of $\Delta_f H^O$. Electronic and zero point energy (E0+ZPE) and relative energy of all isomers are presented in Table 3. The enthalpies of formation of these molecules range from -0.8 (the most stable isomer, tert-butylnitrile) to 52.1(the least stable isomer; 3-dimethyl amino-1-propyne). All values are reported in kcal/mol except electronic and ZPE which are in atomic unit (Hartree/Particle, H/P). Available experimental enthalpies of formation for these isomers are indicated in square brackets in Table 1 and are obtained from the NIST website [18-19].

In column 7 of Table 2, the dipole moments obtained with the G4 composite method for the molecules under consideration are presented. This is necessary because of the large dependence of radio-astronomical observation on the dipole moment of the molecule. The rotational transitions which have been used to detect about 80% of all the known interstellar and circumstellar molecules can only be measured for molecules with permanent dipole moments. The dipole moment also contributes to the intensities of rotational transitions. These intensities scale with the square of the dipole moment, hence, the higher the dipole moment, the higher the intensity of the lines .

**Table 2: $\Delta_f H^O$, structures and dipole moments for $C_5H_9N$ isomers**

| Molecule | Structure | Enthalpy of formation (kcal/mol) | Dipole moment (Debye) | | | |
|---|---|---|---|---|---|---|
| | | | $\mu_A$ | $\mu_B$ | $\mu_C$ | $\mu_{total}$ |
| Tert-butylnitrile | 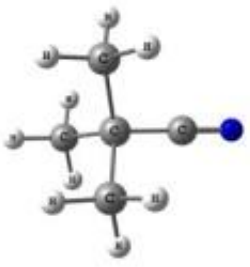 | -0.8[0.8][a,b] | 4.425 | -0.0003 | 0.0003 | 4.425 |
| Isobutyl cyanide | 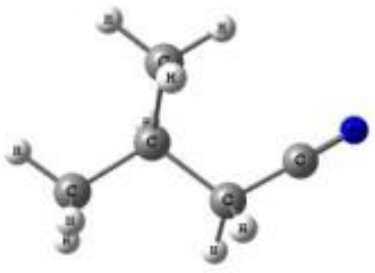 | -0.5 | 4.2128 | - 1.2016 | 0.6351 | 4.4266 |
| n-butyl cyanide | 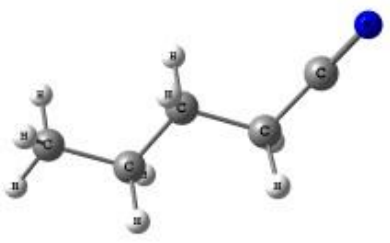 | 0.5 | 4.2309 | 1.8671 | -0.0001 | 4.6245 |
| 2-methylbutanenitrile | 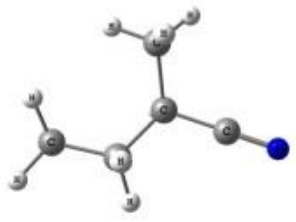 | 0.6[0.6±0.3][a,b] | 4.256 | 1.466 | `0.545 | 4.534 |
| 2-methyl-1-pyrroline | 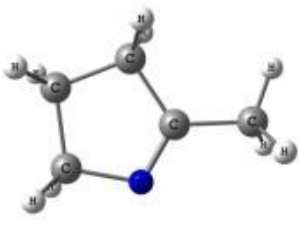 | 1.9 | -0.109 | 2.290 | 0.0001 | 2.292 |
| Cyclopent-3-en-1-amine | 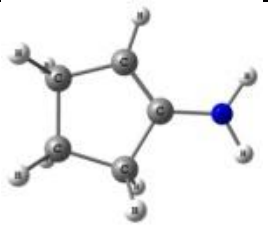 | 7.3 | 1.436 | -0.759 | 0.0001 | 1.625 |
| 2,3,4,5-tetrahydropyridine | 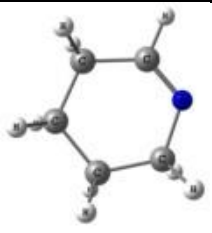 | 15.0 | 1.415 | -1.918 | 0.0001 | 2.384 |
| Tert-butylisonitrile | 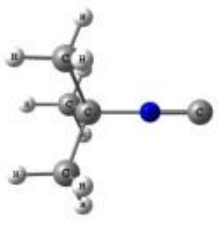 | 17.3 | 3.786 | 0.0001 | 0.0009 | 3.786 |
| 2-isocyanobutane | 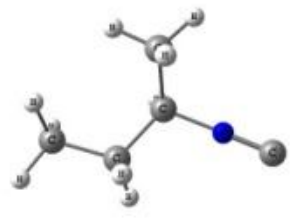 | 19.9 | 3.581 | 1.334 | 0.492 | 3.853 |
| n-butylisocyanide | 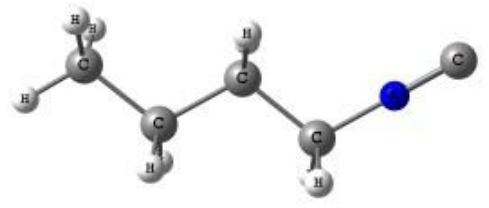 | 21.5 | 3.570 | 1.621 | -0.0001 | 3.921 |
| 1,2,3,6-tetrahydropyridine | 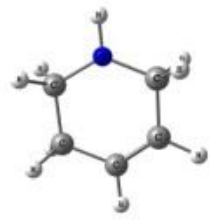 | 24.7 | 0.120 | -0.345 | 0.000 | 0.366 |
| 3-methylene cyclobutanamine | 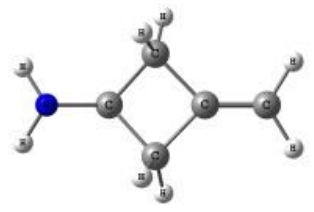 | 32.6 | 1.019 | -0.0002 | 1.312 | 1.661 |
| 2-methyl-3-butyn-2-amine | 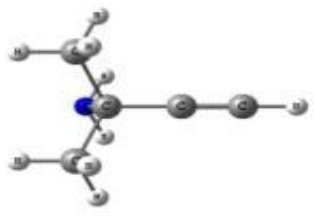 | 39.7 | 0.694 | 0.001 | -0.016 | 0.695 |
| 4-pentyn-1-amine | 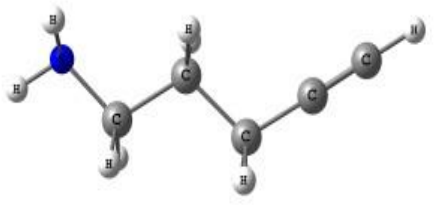 | 43.7 | 1.065 | 0.950 | 1.116 | 1.811 |
| 1-ethyl-2-methyleneazaridine | 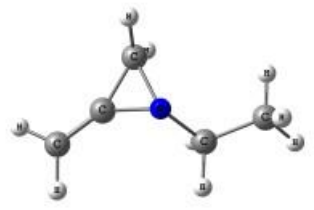 | 44.6 | -0.750 | 0.272 | 1.191 | 1.434 |
| 3-dimethyl amino-1-propyne | 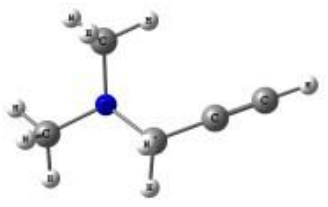 | 52.1 | -0.759 | -0.075 | 0.759 | 1.077 |

[a]Hall and Baldt[6], [b]NIST[7]

**Table 3: Electronic + zero point energy (E0+ZPE) and relative energy for $C_5H_9N$ isomers by using G4 method**

| Molecule | Electronic (E0) + zero point energy (ZPE) (H/P) | Relative energy (kcal/mol) |
|---|---|---|
| Tert-butylnitrile | -250.585838 | 0 |
| Isobutyl cyanide | -250.585182 | 0.411646 |
| n-butyl cyanide | -250.583645 | 1.376128 |
| 2-methylbutanenitrile | -250.583258 | 1.618974 |
| 2-methyl-1-pyrroline | -250.581338 | 2.823793 |
| Cyclopent-3-en-1-amine | -250.558466 | 17.17619 |
| 2,3,4,5-tetrahydropyridine | -250.571024 | 9.295925 |
| Tert-butylisonitrile | -250.557080 | 18.04592 |
| 2-isocyanobutane | -250.552865 | 20.69087 |
| n-butylisocyanide | -250.550339 | 22.27596 |
| 1,2,3,6-tetrahydropyridine | -250.556476 | 18.42493 |
| 3-methylene cyclobutanamine | -250.531658 | 33.58682 |
| 2-methyl-3-butyn-2-amine | -250.518612 | 42.18495 |
| 4-pentyn-1-amine | -250.515075 | 44.40445 |
| 1-ethyl-2-methyleneazaridine | -250.512984 | 45.71657 |
| 3-dimethyl amino-1-propyne | -250.501462 | 52.94674 |

Rotational excited states can be populated at 10-100 K and decay by spontaneous emission, infrared emission can be observed if molecules are occasionally excited by high energy photons emitted by hot stars in the vicinity of the cloud which is a rare event. Interestingly, all the molecules considered here have non-zero permanent electric dipole moment ranging from 0.366 (for 1,2,3,6-tetrahydropyridine) to 4.628 Debye (for n-butyl cyanide) which implies that their rotational spectra can be measured, thus providing the necessary tool for their astronomical searches.

The enthalpies of formation, structure and the current astronomical status for the $C_4H_7N$ isomers are presented in Table 1. The two astronomically observed isomers of this group are the most stable isomers (with the least E0+ZPE and enthalpies of formation) as compared to other isomers of the group. From the results of the high-level quantum chemical calculations performed on the $C_5H_9N$ isomers presented in Table 2, it is crystal clear that the most stable isomer of the group is tert-butylnitrile, a branched chain molecule with a dipole moment of 4.425 Debye and a negative enthalpy of formation (-0.8 kcal/mol). The enthalpies of formation estimated with the G4 composite method are in good agreement with the experimentally measured values as shown in both tables [18]. The microwave spectrum of this molecule was measured **in 1962** [28]. From the measured rotational constants, more rest frequencies in the range of interest for astronomical search can be predicted. The astronomical searches of this isomer is yet to be reported anywhere in literature to the best of our knowledge. However, it remains the best candidate for astronomical observation among all the isomers of the $C_5H_9N$ isomeric group being the most stable isomer of the group thus it could be detected in the interstellar or circumstellar medium. The second most stable isomer of the $C_5H_9N$ isomeric group is isobutyl cyanide, an important intermediate in the pharmaceutical industry and a key starting material in the production of diazinon, an organo-phosphorus pesticide [29]. It has a dipole moment of 4.427 Debye similar to that of tert-butylnitrile. The rotational frequencies of this molecule are not known and as such there has not been any report of its astronomical search. With its stability and high dipole moment, it remains a potential candidate for astronomical observation pending the availability of accurate rotational frequencies.

The third most stable isomer of the $C_5H_9N$ isomeric group is n-butyl cyanide, the next largest unbranched alkyl cyanide after n-propyl cyanide that was detected in the interstellar medium **in 2009** [16]. The rotational spectrum of this molecule in the range of 5 to 22 GHz has been measured [30]. This measurement has been extended to the shorter millimeter wavelengths to aid its astronomical searches. The radio-astronomical searches for n-butyl cyanide in the Sagittarius B2(N) molecular cloud yielded only upper limits of its column density without any successful detection [31]. This is the only isomer of the $C_5H_9N$ isomeric group that has been astronomically searched for though it is not the most stable molecule of the group. The upper limit obtained for n-butyl cyanide points to the possibility of a successful detection for the most stable isomer, tert-butylnitrile (branched chain molecule) which is more stable than n-butyl cyanide.

The next most stable isomer after the linear isomer, n-butyl cyanide is 2-methylbutanenitrile which is very similar to the linear isomer in terms of energy and dipole moment. However, the rotational transitions of this molecule which could warrant its astronomical searches are not known. There is a drastic decrease in stability and dipole moment of the other isomers after 2-methylbutanenitrile thus making their observations less feasible. The high stability of the cyanides than their corresponding isocyanides could be seen here. For example, tert-butylnitrile (-0.8 kcal/mol) being more stable than tert-butylisonitrile (17.3 kcal/mol); n-butyronitrile (0.5 kcal/mol) being more stable than n-butylisocyanide (19.9 kcal/mol), etc. The cyclic isomers are in most cases found to be less stable than their corresponding straight or branched chain counterparts in majority of the groups, resulting in few numbers of known cyclic molecules in space. This is also observed here.

## 3.1 Conformers

Spectral intensity dilution may happen due to the populations of conformers. Thus, before proposing to observe any particular species in the interstellar region, it is essential to know the relative energies between all its and their conformers and also to point out the lowest energy conformer of the species of interest [32-33]. Based on our calculations (enthalpy of formation) it is evident that Tert-butylnitrile, Isobutyl cyanide and n-butyl cyanide are the three most stable isomers of the $C_5H_9N$ isomeric group. Thus, in the following, we have discussed about the conformers of these three candidates.

*Tert-butylnitrile:*

We found that Tert-butylnitrile does not have any conformers.

*n-butyl cyanide:*

[34] Crowder et al. (1989) stated that n-butyl cyanide could have four conformers. Their gas phase infrared spectrum showed that GA conformer is the most abundant (having 46%). According to their experiment, the next three conformers are AA (30%), AG (13%) and GG' (11%). According [31], n-butyl cyanide could exist into five conformers. Ordu et al. (2012) and references therein showed that AA is the most stable conformer and GG being the highest energy conformer. Our computations with the G4 composite method also found that AA is the most stable conformer and GG is the highest energy conformer. Our calculated values follow the same trend as mentioned in [31] except AG and GA. We found that AG has slightly lower energy in comparison to GA whereas [31] pointed the slightly opposite trend between these two conformers. Details of this conformer analysis are presented in Table 4.

**Table 4: Relative energies and dipole moment components of various conformers of n-butyl cyanide**

| Conformers | E0+ZPE (Hartree/p) | ZPE (Hartree/p) | Relative energy (kcal/mol) | Dipole moments | | | |
|---|---|---|---|---|---|---|---|
| | | | | $\mu_a$ | $\mu_b$ | $\mu_c$ | $\mu_{Total}$ |
| **AA (Anti-Anti)** | **-250.583645** | **0.131108** | **0** | **4.2309** | **1.8671** | **-0.0001** | **4.6245** |
| **AG (Anti-Gauche)** | **-250.583642** | **0.131111** | **0.001883** | **4.5519** | **-0.2170** | **0.9892** | **4.6632** |
| **GA (Gauche-Anti)** | **-250.583516** | **0.131173** | **0.080949** | **-2.8906** | **3.1885** | **-0.5005** | **4.3328** |
| **GG' (Gauche-Gauche')** | **-250.582253** | **0.131286** | **0.873493** | **3.1873** | **2.7352** | **-1.0501** | **4.3293** |
| **GG (Gauche-Gauche)** | **-250.580681** | **0.131239** | **1.859938** | **2.8191** | **-3.2394** | **-0.5640** | **4.3312** |

*Isobutyl cyanide:*

[29] discussed various conformers of Isobutyl cyanide. According to their study, TG conformer is found to be the lowest energy conformer whereas GG conformer is found to be the highest energy conformer. From our calculations with the G4 composite method, we found the similar trend. However, we have noticed that TG and GT conformers are iso-energetic. Relative energy values including zero point corrections and dipole moment components of the various conformers of Isobutyl cyanide are presented in Table 5.

**Table 5: Relative energies and dipole moment components of various conformers of isobutyl cyanide**

| Conformers | E0+ZPE (Hartree/p) | ZPE (Hartree/p) | Relative energy (kcal/mol) | Dipole moments | | | |
|---|---|---|---|---|---|---|---|
| | | | | $\mu_a$ | $\mu_b$ | $\mu_c$ | $\mu_{Total}$ |
| TG (Trans-Gauche) | -250.585182 | 0.130633 | 0.00 | 4.2108 | - 1.2011 | 0.6341 | 4.4244 |
| GT (Gauche-Trans) | -250.585182 | 0.130634 | 0.00 | -4.2108 | -1.2005 | 0.6357 | 4.4245 |
| GG (Gauche-Gauche) | -250.584845 | 0.130708 | 0.211471 | -3.5459 | 0.0000 | 2.5386 | 4.3609 |

### 3.2 Rotational Intensity:

Charnley et al., [35] pointed out that for an optically thin emission, an idea about the antenna temperature could be made by calculating the intensity of a given rotational transition. This intensity is proportional to $\mu^2/Q(T_{rot})$ , where μ is the electric dipole moment and $Q(T_{rot})$ is the partition function at rotational temperature ($T_{rot}$) and calculated by using the following relation

$$Q(T_{rot}) = 5.3311 \times 10^6 \sqrt{(T^3/ABC)} \quad \text{equ. 1}$$

where, A, B and C are the rotational constants shown in Table 6. Our calculated constants are in good agreement with the existing values. For example, in case of n-butyl cyanide, our calculated values of A, B and C are 15300.020, 1319.670 and 1252.401 MHz respectively whereas [31] found 15028.687, 1334.106 and 1263.857 MHz respectively for A, B and C. These imply the percentage deviations of 1.8, 1.1 and 0.9% respectively for the A, B and C rotational constants. In Fig. 1, we compare the expected intensities for all the species with respect to tert-butylnitrile (the most stable isomer of this group). For simplicity, here we assume that all these isomers are having same abundances. Here for the calculation, we use the best fit LTE parameters obtained by [11] for the observation of n-butyl cyanide. Thus, we adopted a rotational temperature of 150K for our calculation. From Fig. 1, it is clear that tert-butylnitrile would have the strongest transition in between the isomers of this isomeric group. This is in agreement with the ground state energy and enthalpies of formation as discussed earlier.

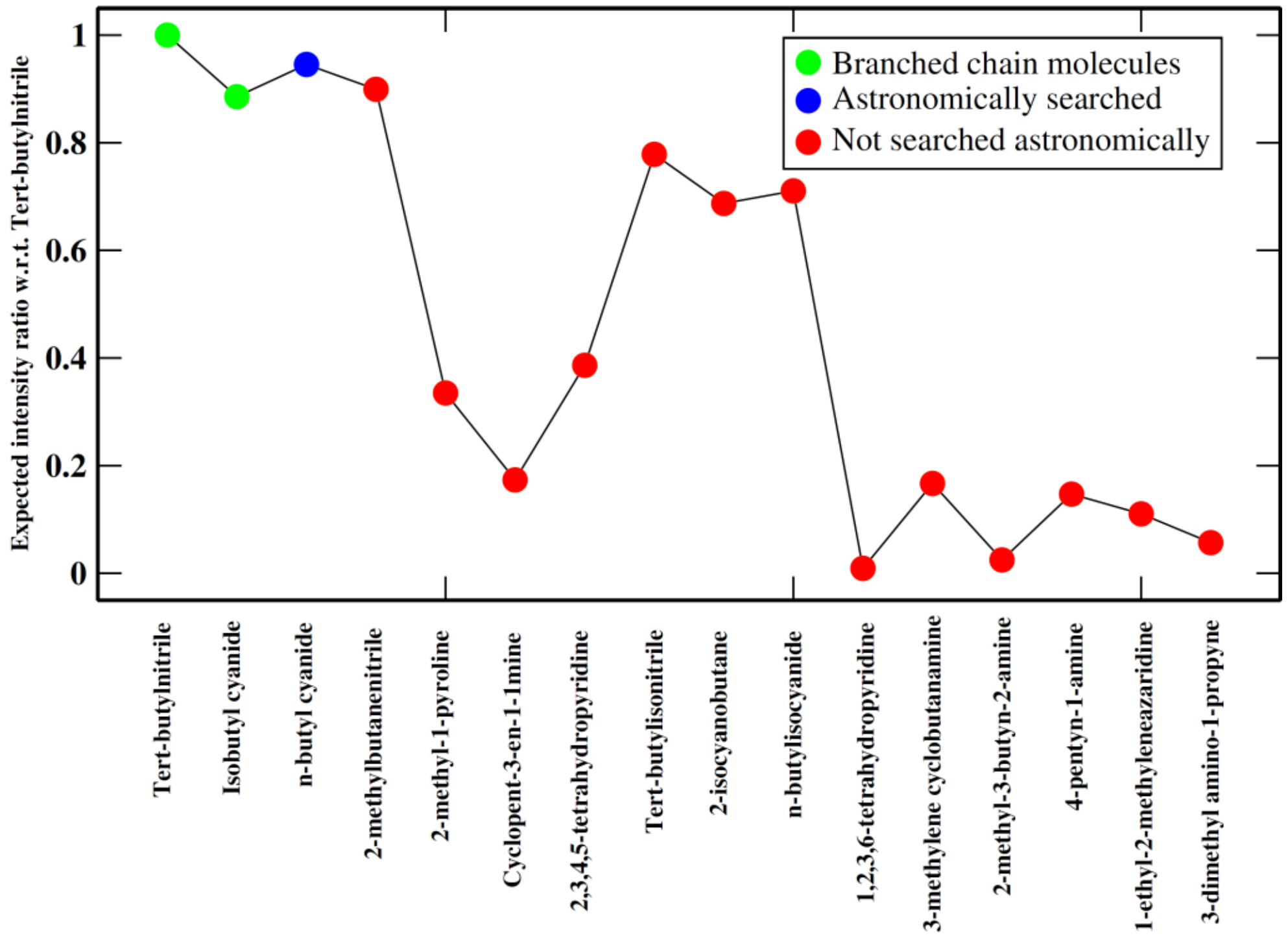

Fig.1: Expected intensity ratio by assuming the same column density and rotational temperature

According to this expected intensity ratio, the second most probable candidate for astronomical detection appears to be n-butyl cyanide which is in contradiction to the ground state energy and enthalpy of formation sequence. Isobutyl cyanide which stood second in the sequence of ground state energy and enthalpy of formation appears to be third as per the expected intensity ratio. Thus, based on our ground state energy and enthalpy of formation sequence and expected intensity ratio, tert-butylnitrile is found to be the most probable candidate for astronomical detection in the $C_5H_9N$ isomeric group. In order to firmly detect this molecule in the ISM, spectral intensities, along with the frequencies for rotational transitions of tert-butyl cyanide is predicted by using experimentally calculated rotational constants [36] and experimentally obtained dipole moment [37] followed by the SPCAT program [38]. This catalog file is prepared in JPL format and is given as supplementary material with this article.

**Table 6: Rotational and quartic centrifugal -distortion constants values of $C_5H_9N$ isomers calculated by using B3LYP/ 6-311G++(d,p) method.**

| Species | Rotational constants | Values in MHz | Quartic centrifugal-distortion constants | Values in KHz |
|---|---|---|---|---|
| Tert-butylnitrile | A | 4485.109 | $\Delta_J$ | 0.4507 |
| | B | 2747.149 | $\Delta_K$ | 3.289 |
| | C | 2747.077 | $\Delta_{JK}$ | 0.1368 |
| | A | --- | $\delta_J$ | 0.0258 |
| | B | (4600.00)[a] | $\delta_K$ | -647.44 |
| | C | (2749.909)[a] | | |
| | | --- | | |
| Isobutyl cyanide | A | 7319.902 | $\Delta_J$ | 0.3233 |
| | B | 2079.579 | $\Delta_K$ | 8.037 |
| | C | 1740.880 | $\Delta_{JK}$ | 3.442 |
| | | | $\delta_J$ | 0.0649 |
| | | | $\delta_K$ | -1.791 |
| n-butyl cyanide | A | 15300.020 | $\Delta_J$ | 0.0797 |
| | B | 1319.670 | $\Delta_K$ | 152.82 |
| | C | 1252.401 | $\Delta_{JK}$ | 21.11 |
| | A | (15028.687)[b] | $\delta_J$ | -0.0059 |
| | B | (1334.106)[b] | $\delta_K$ | 0.0583 |
| | C | (1263.857)[b] | | |
| 2-methylbutanenitrile | A | 6424.296 | $\Delta_J$ | 0.1475 |
| | B | 2201.255 | $\Delta_K$ | 14.34 |
| | C | 1756.028 | $\Delta_{JK}$ | 4.262 |
| | | | $\delta_J$ | 0.0138 |
| | | | $\delta_K$ | 1.344 |
| 2-methyl-1-pyrroline | A | 7133.547 | $\Delta_J$ | 0.2353 |
| | B | 3168.904 | $\Delta_K$ | 0.7124 |
| | C | 2334.006 | $\Delta_{JK}$ | 1.298 |
| | | | $\delta_J$ | 0.0401 |
| | | | $\delta_K$ | 0.5307 |
| Cyclopent-3-en-1-amine | A | 5955.298 | $\Delta_J$ | 1.540 |
| | B | 3532.729 | $\Delta_K$ | 1.537 |
| | C | 2660.970 | $\Delta_{JK}$ | 0.4308 |
| | | | $\delta_J$ | 0.1740 |
| | | | $\delta_K$ | 2.386 |
| 2,3,4,5-tetrahydropyridine | A | 4870.596 | $\Delta_J$ | 1.142 |
| | B | 4682.369 | $\Delta_K$ | 0.9346 |
| | C | 26322.255 | $\Delta_{JK}$ | -1.876 |
| | | | $\delta_J$ | -0.0070 |
| | | | $\delta_K$ | -3.157 |
| Tert-butylisonitrile | A | 4496.664 | $\Delta_J$ | 0.5213 |
| | B | 2917.748 | $\Delta_K$ | 3.117 |
| | C | 2917.638 | $\Delta_{JK}$ | -0.020 |
| | A | -- | $\delta_J$ | -0.033 |
| | B | (2932.1759)[c] | $\delta_K$ | -542.93 |
| | C | -- | | |
| 2-isocyanobutane | A | 6521.507 | $\Delta_J$ | 0.1588 |
| | B | 2319.387 | $\Delta_K$ | 14.10 |
| | C | 1838.748 | $\Delta_{JK}$ | 4.363 |
| | | | $\delta_J$ | 0.014 |
| | | | $\delta_K$ | 1.377 |
| n-butylisocyanide | A | 15418.749 | $\Delta_J$ | 0.082 |
| | B | 1377.132 | $\Delta_K$ | 146.06 |
| | C | 1305.007 | $\Delta_{JK}$ | 20.64 |
| | | | $\delta_J$ | -0.0059 |
| | | | $\delta_K$ | 5.74 |
| 1,2,3,6-tetrahydropyridine | A | 4931.777 | $\Delta_J$ | 1.053 |
| | B | 4741.531 | $\Delta_K$ | 0.039 |
| | C | 2631.568 | $\Delta_{JK}$ | -0.9817 |
| | | | $\delta_J$ | 0.046 |
| | | | $\delta_K$ | 2.837 |
| 3-methylene cyclobutanamine | A | 9712.786 | $\Delta_J$ | 0.093 |
| | B | 2386.903 | $\Delta_K$ | 19.66 |
| | C | 2049.693 | $\Delta_{JK}$ | 4.323 |
| | | | $\delta_J$ | -0.010 |
| | | | $\delta_K$ | -0.017 |
| 2-methyl-3-butyn-2-amine | A | 4647.936 | $\Delta_J$ | 0.443 |
| | B | 2722.380 | $\Delta_K$ | 2.999 |
| | C | 2699.013 | $\Delta_{JK}$ | 0.406 |
| | | | $\delta_J$ | -0.030 |
| | | | $\delta_K$ | 13.51 |
| 4-pentyn-1-amine | A<br>B<br>C | 15867.178<br>1317.724<br>1248.851 | $\Delta_J$<br>$\Delta_K$<br>$\Delta_{JK}$<br>$\delta_J$<br>$\delta_K$ | 0.090<br>225.87<br>19.00<br>0.029<br>-0.513 |
| 1-ethyl-2-methyleneazaridine | A<br>B<br>C | 8810.985<br>2225.940<br>1904.983 | $\Delta_J$<br>$\Delta_K$<br>$\Delta_{JK}$<br>$\delta_J$<br>$\delta_K$ | 0.2496<br>36.05<br>5.855<br>0.0044<br>1.78 |
| 3-dimethyl amino-1-propyne | A<br>B<br>C | 8081.087<br>2143.978<br>1806.939 | $\Delta_J$<br>$\Delta_K$<br>$\Delta_{JK}$<br>$\delta_J$<br>$\delta_K$ | 0.2759<br>27.72<br>6.404<br>0.094<br>2.933 |

[a] Kisiel, [36],[b] Ordu et al., [31], [c] Howard [39]

## 4. LTE modeling

We have used the following input parameters for the LTE modeling to mimic a typical high mass star forming region [40].

Column density of $H_2$=$10^{24}$ $cm^{-2}$
Excitation temperature ($T_{ex}$) =100K
FWHM=5 km/s
$V_{LSR}$=74 km/s
Source size=3”
Abundance of Tert-butylnitrile = $10^{-11}$ with respect to $H_2$

For this modeling, we have used the frequency file generated by SPCAT programme with experimentally obtained rotational constants (A= 4600.00 MHz, B= 2749.90 MHz, C= 2749.90 MHz from Kisiel, 1985) and use CASSIS software. Since the abundance of this complex molecule is expected to be low, it is essential to use the high-spatial and high spectral resolution observations from the ALMA. Specifically, we can use the lower frequency bands of ALMA. i.e., Bands 1,2 and 3 (31-116 GHz). Currently ALMA band 3 is in operation while bands 1 and 2 are yet to be available. From the following table, we can have the strongest line (intensity ~25 mK) in Band 3 around 115 GHz. Using Band 3, we can detect this transition approximately 19 hours of integration time by assuming a spectral resolution of 1 km/s, a sensitivity of 8.33 mK (a signal-to-noise ratio of 3), an angular resolution 3 arcseconds and 40 antennas of 12 meter array in the ALMA time estimator.

**Table 7: Calculated line parameters in the millimetre wave region for tert-butyl cyanide with ALMA**

| Species | ALMA | Frequency | $J_{Ka'Kc'}$- | Eup | Aij | Intensity |
|---|---|---|---|---|---|---|

| | Band | (GHz) | $J_{ka''kc''}$ | | | (mK) |
|---|---|---|---|---|---|---|
| | Band 1 | 43.99274<br>43.99257 | $8_{3\,6}$-$7_{3\,5}$<br>$8_{3\,5}$-$7_{3\,4}$ | 10.30<br>10.24 | $3.42\times10^{-5}$<br>$3.37\times10^{-5}$ | 2.55<br>2.35 |
| Tert-butyl cyanide | Band 2/ Band 3 | 88.01920<br>88.01922 | $16_{4\,13}$ - $15_{4\,12}$<br>$16_{4\,12}$ - $15_{4\,11}$ | 37.32<br>37.32 | $1.15\times10^{-9}$<br>$1.15\times10^{-9}$ | 17<br>17 |
| | Band 3 | 115.54571<br>115.54559 | $21_{4\,17}$ - $20_{4\,16}$<br>$21_{4\,18}$ - $20_{4\,17}$ | 62.41<br>62.41 | $8.24\times10^{-10}$<br>$8.24\times10^{-10}$ | 25<br>25 |

Band 1 = 31-45 GHz
Band 2= 67-90 GHz
Band 3= 84-116 GHz

## 5. Conclusion

The aim of this work has been in examining the best candidate for astronomical observations among the $C_5H_9N$ isomers by searching for the stability of all isomers. From the results of our high-level quantum chemical calculations, the most stable isomer of the $C_5H_9N$ isomeric group is tert-butylnitrile, a branched chain molecule followed by Isobutyl cyanide, another branched chain molecule. The only isomer of this group that has been astronomically searched for with only upper limit of column density determined is the linear isomer; n-butyl cyanide which ranks third on the energy scale among other isomers of the group; thus, not the most stable isomer of the group in the interstellar medium. In accordance with the optimized energy, relative energy, enthalpies of formation and calculation regarding expected intensity ratio, tert-butylnitrile is found to be most probable candidate for astronomical detection. Thus, we are proposing a possible successful astronomical observation of tert-butylnitrile around the same region where n-butyl cyanide was attempted. Spectroscopic parameters and dipole moments that could guide the astronomical searches for these molecular species have been provided in the study**.**

**Acknowledgement:** EEE acknowledges a research fellowship from the Indian Institute of Science, Bangalore. AD & PG are grateful to ISRO (Grant No.ISRO/RES/2/402/16-17) for the partial financial support.